\documentclass[12pt]{article}
\usepackage[utf8]{inputenc}
\usepackage[english]{babel}
\usepackage[a4paper,lmargin={3cm},rmargin={3cm},tmargin={3cm},bmargin={3cm}]{geometry}
\usepackage{mathptmx}
\usepackage{amsmath}
\usepackage{color}
\usepackage{hyperref}
\usepackage{float}
\usepackage[super]{nth}
\usepackage[symbol]{footmisc}
\usepackage{booktabs}
\usepackage{longtable}
\usepackage{lscape}
\usepackage{tabularx}
\usepackage{multirow}
\usepackage{threeparttable}
\usepackage{dcolumn}
\usepackage{graphicx}
\usepackage[export]{adjustbox}
\usepackage{subcaption}
\usepackage[justification=centering,font=small,labelfont=bf]{caption}
\usepackage{dcolumn}
\usepackage{epstopdf}
\usepackage{fancyhdr}
\usepackage{amsfonts}
\usepackage{amsmath}
\usepackage{epsfig}

\begin{document}

\title{Reframing the S\&P500 Network of Stocks along the \nth{21} Century}
\author{Tanya Ara\'{u}jo\thanks{\emph{Corresponding author} (tanya@iseg.utl.pt) } \hspace{0.2cm} and Maximilian Göbel\\
ISEG, University of Lisbon and UECE\\ R. Miguel Lupi 20, 1249-078 Lisboa, Portugal}
\date{}
\maketitle

\begin{abstract}
Since the beginning of the new millennium, stock markets went through every state from long-time troughs, trade suspensions to all-time highs. The literature on asset pricing hence assumes random processes to be underlying the movement of stock returns. Observed procyclicality and time-varying correlation of stock returns tried to give the apparently random behavior some sort of structure. However, common misperceptions about the co-movement of asset prices in the years preceding the \emph{Great Recession} and the \emph{Global Commodity Crisis}, is said to have even fueled the crisis' economic impact. Here we show how a varying macroeconomic environment influences stocks' clustering into communities. From a sample of 296 stocks of the S\&P 500 index, distinct periods in between 2004 and 2011 are used to develop networks of stocks. The Minimal Spanning Tree analysis of those time-varying networks of stocks demonstrates that the crises of 2007-2008 and 2010-2011 drove the market to clustered community structures in both periods, helping to restore the stock market's ceased order of the pre-crises era. However, a comparison of the emergent clusters with the \textit{General Industry Classification Standard} conveys the impression that industry sectors do not play a major role in that order.
\end{abstract}
\textbf{Keywords:} S\&P500, Network Analysis, Minimal Spanning Trees, Industrial Clusters, Great Recession, Global Commodity Crisis, Community Detection.

\section{Introduction}
Since the beginning of the \nth{21} century, financial markets have undergone turbulent times. Especially stock markets have experienced several ups and downs, while successively breaking through all-time highs. The S\&P 500 Index lost almost 50\% within months after the bursting of the so-called "dot-com" bubble in March 2000, when it reached its first trough of the millennium at 776.76 points on 10/09/2002. It took the index almost exactly five years, until it had fully recaptured its peak value of the year 2000, when it closed at 1529.03 points on 09/19/2007. The bullish sentiment persisted another few weeks until the index peaked on 10/09/2007 - exactly five years after its previous trough - at 1565.15 points. However, rumors about falling housing prices, precedingly lax lending standards and a fragil financial system made the S\&P 500 turn. For the following 18 months, the bears took over: the S\&P 500 fell to its lowest value since the beginning of the millennium on 03/09/2009 at 676.53 points. Similar to the period after the "dot-com" crisis, the index had lost more than 56\% of its previous turning-point value on 03/09/2009. The difference to the events at the beginning of the new century was, however, the time frame, as the crash of the years 2007 through 2009 had occurred within only 18 instead of 30 months. Nevertheless, from March 2009 on, bearish sentiment seems to have vanished, with not only the S\&P 500, but stock indices all around the world ever rising. Between its trough in spring 2009 and the end of 2015, the S\&P 500 more than trippled its score. The bullish sentiment even persisted until autumn 2018, when the index reached its all-time high, ranking at about 4.3 times above its value on 03/09/2009.

A complete assessment of the forces driving these fluctuations during the first decade of this century and the consistently bullish behavior thereafter, demands the merging of economic, mathematical and psychological sciences. The literature on the behavior of asset prices, however, generally assumes random processes to be the underlying causes for the dynamics of stock returns \cite{Man99}. A vast literature of models, starting with \cite{FF93}, tried to identify common factors of single stock returns and to shed further light on the seemingly random behavior of stock returns.

Nevertheless, one commonly agreed perception about stock prices is their procyclicality (\cite{Co97}, \cite{Ko96}, \cite{MePr03}). Thus, the correlation of stock returns plays a major role in portfolio construction and financial modelling. As Marvin (\cite{Mar15}) points out, these correlations of the first difference of one-day logged stock prices are, however, not constant over time and may even reverse during times of crisis. The widespread misperception of the market dynamics by investors and financial modellers in the beginning of the 2000s is said to have amplified the severity of the critical period 2007-2008, i.e., the \textit{Great Recession} (\cite{Poz10}).

Shortly after, the two-year period of 2010 and 2011 was also characterized by both a steep in- as well as decrease in global commodity prices.
The \textit{All Commodities Price Index} did not surpass its peak of 2008, but the sub-index of \textit{Non-Fuels} experienced a sharp increase in late 2010 before crushing down from mid 2011 onwards. The literature identifies both demand side as well as supply side effects to be accountable for the rally in commodity prices. Hochman and co-authors (\cite{Hoch14}) name speculation, rising energy prices and low agricultural productivity growth, but above all a depreciating U.S. dollar as macroeconomic reasons for the run up in food commodity prices in late 2010. According to an IMF report (\cite{IMF12}), the rise in food prices was mostly caused by poor harvests in Asia and Russia, whereas the decline in 2011 is mostly attributable to a slowing down of global economic growth.

Assessing the co-movements of stock returns and time-varying market structures are, hence, key to understanding and reacting to the effects of a changing macroeconomic environment. The purpose of this paper is to shed further light on the community structure of the S\&P 500 stocks in between the years 2004 through 2011, by emphasizing the roles of both the \textit{Great Recession} and the \emph{Global Commodity Crisis}. Facilitating the understanding of the dynamics of community structure over time and especially during turbulent periods improves the knowledge on the role of some macroeconomic variables and their influence on stock returns.

The way stocks aggregate into either clusters of industry sectors or into clusters defined by the simultaneous co-movement of stock returns is envisioned to contribute to uncover the dynamics of the S\&P 500 stock market.
Clustering in networks, where links are based on simultaneous co-movements of stock returns, is a measure of synchronization in the market. As such, clustering may provide information independent of other global market indicators, improving the search for {\it economic factors} which may be neither industry sectors nor other obvious economic facts (\cite{Ara00}-\cite{El14}).

In this setting, our approach is three-fold: at first, networks of stocks are induced from data of both the \emph{business-as-usual} period of 2004-2005, and two critical periods: the \textit{Great Recession} in 2007-2008 and the \emph{Global Commodity Crisis} in 2010-2011. Their Minimal Spanning Trees (MST) are computed to filter out the strongest links among companies based on the Euclidean distance between the one-day logged differences of stock prices. The second stage comprises the use of \textit{Gephi's} Community Detection Algorithm, based on Blondel (\cite{Bl08}), to identify community formation in the MST representation of the networks. Each MST is also characterized by the diameter ($d$), the characteristic path length ($C$) and the maximum degree ($mk$). The last stage compares the resulting partition with a "natural" classification, namely the one that is defined by the \textit{Global Industry Classification Standard} (GICS), separating the 296 stocks into 11 distinct industry sectors.

The remainder of the paper is structured as follows: Section 2 shortly describes the data set used for the forthcoming analysis, before a short summary of network induction and modularity detection techniques is given in Section 3 in order to facilitate their interpretation in Section 4. The last section concludes.

\section{Data} \label{Data}

The underlying data set comprises 296 companies of the \textit{Standard\&Poor's 500} index, covering the period from 01/03/2000 to 12/31/2015. Daily stock prices have been extracted from the information network Bloomberg \cite{Bloom16}. The analyzed companies have thus survived the first 15 years of the \nth{21} century and have, in addition, not been delisted from the S\&P 500 index.  The community structure of the S\&P 500 was examined in three distinct two-year periods: the pre-crises period (2004-2005), and the turbulent periods of the \emph{Great Recession} (2007-2009) and the \emph{Global Commodity Crisis }(2010-2011).
For further comparison and differentiation of companies, their daily market capitalization served as a proxy for a company's size.\footnote{Market capitalization is defined as the total current market value of all of a company's outstanding shares, as stated by the data provider. A company's daily market capitalization is the product of the number of \textit{current shares outstanding} and the trading day's \textit{close-of-business stock price.}}.

\section{Method} \label{sec:Method}

Network approaches have been a common practice in the analysis of systems, whose main focus relies on a relational nature.
It is a tool used frequently in the studies of financial systems and especially in the analysis of the dynamics of market stocks (\cite{Man99},\cite{Ara00}-\cite{El14}).

\subsection{Networks of stocks} \label{sec:NetworksofStocks}

The adoption of networks has often been based on the notion of {\it distance}.
Depending on the circumstances, distance may be measured by the strength of
interaction between the agents of a system, by their spatial distance or by some other
criterion expressing the existence of a link between the agents.

Based on the notion of distance, global and local parameters have been defined to characterize the
connectivity structure of the induced networks. Here, we are mostly interested in three global parameters:
the {\it modularity} measure ($Q$), the characteristic path length ($C$) and the network diameter ($d$).

Correlation-based metrics are frequently used for computing metric-compliant distance measures.
We follow Mantegna (\cite{Man99}) in computing the distance between any two stocks as:

\begin{align}
d(i,j) = \sqrt{2(1 - \rho_{ij})} \label{eq4} \
\end{align}
where $\rho_{ij}$ is the Pearson Correlation Coefficient computed for each pair of stock returns $(r_{i},r_{j})$, which are derived from the one-day log differences of $n$ stock prices recorded in time-series of length $T$:

\begin{align}
r_{it} = log \left( \frac{p_{i,t}}{p_{i,t-1}} \right) \qquad ,
\end{align}
- $p_{i,t}$ being the stock price of company $i$ at time $t$.

The methods for describing the way community structures and its associated modularity measure are computed can be found in the literature under the notion of hierarchical clustering (\cite{New06},\cite{NewGi04}). Stock return distances are then used to induce networks of stocks, where parameters can be measured in order to characterize the network structure. For this purpose, we construct a graph from the weight $w_{ij}$, which measures the inverse distance ($d_{ij}=\mid 1/w_{ij}\mid $) of each pair of stock returns $(r_{i},r_{j})$ over a certain time horizon $T$. Such a distance-based measure, $w_{ij}$, corresponds to the connection strength between stocks $i$ and $j$. In so doing, the resulting network of stocks, $N$, is a complete, undirected and weighted network.

However, the computation of global and local parameters usually applies to graph structures that are
sparse. Since the networks we work with are fully-connected structures, a
first step is targeted at obtaining a sparse representation of the network,
with the {\it degree of sparseness} generated endogenously,
instead of an {\it a priori} specification. When looking for a
suitable degree of sparseness, disconnectivity shall be avoided.

The  Minimal Spanning Tree (MST) is a representation of a network, where sparseness replaces full-connectivity in a suitable way.

\subsection{Minimal Spanning Tree}

From the $n\times n$ distance matrix a hierarchical clustering is then performed using the {\it nearest
neighbor method}. Initially, $n$ clusters, corresponding to the $n$ agents, are
considered. Then, at each step, two clusters $c_{i}$ and $c_{j}$ are clumped
into a single cluster if
\[
d_{c_{i}c_{j}}=\min \left\{ d_{c_{i}c_{j}}\right\}
\]
with the distance between clusters being defined by
\[
d_{c_{i}c_{j}}=\min \left\{ d_{pq}\right\}
\]
with $p\in c_{i}$ and $q\in c_{j}$

This process continues until a single cluster remains. This clustering
algorithm is also known as the {\it single link method}, being the method by
which one obtains the Minimal Spanning Tree (MST) of a graph. In a connected
graph, the MST is a tree of $n-1$ edges that minimizes the sum of the edge
distances.

In a network with $n$ agents, the hierarchical clustering process takes $n-1$
steps to be completed, and uses, at each step, a particular distance $%
d_{ij}\in D_{W}$ to clump two clusters into a single one.

Linking the nodes with the lowest distance (highest strength) allows to efficiently assess the intensity of connections between stocks
and between different industry sectors within a given portfolio.

\subsection{Modularity and communities} \label{mod}

Examining the community structure of the S\&P 500 in three distinct two-year periods, this paper builds up on the community structure detection approach by Newman (\cite{New06}), which defines the modularity, $Q$, as the difference in the number of edges within a cluster and the number of edges in a random network. The crucial assumption is based on the fact that a random network does not exhibit any community structure (\cite{Iso15}), whereas a high modularity suggests a large deviation of the detected clustering from a completely randomized network.

\bigskip

\textbf{{\textit{Gephi's} modularity measure}}

\bigskip

 \textit{Gephi's} measure of modularity, $Q^G$, is based on Blondel (\cite{Bl08}). The authors extended Newman's (\cite{New06}) initial algorithm by improving the efficiency of computation and the quick reduction of the number of communities.\footnote{Furthermore, the \textit{resolution limit problem}, which Fortunato \& Barthélemy (\cite{Fo07}) and Isogai (\cite{Iso15}) describe as Newman's (\cite{New06}) struggle to detect small clusters, is mitigated by Blondel (\cite{Bl08}) algorithm.}

As presented earlier, our method is three-fold: at first, networks of stocks are induced and their corresponding minimal spanning trees are computed. Next, we make use of \textit{Gephi's} Community Detection Algorithm to identify communities in the MST representation of the network. Then, each MST is also characterized by their diameter and characteristic path length. The last stage compares the resulting partition in communities with the partition of stocks defined by their industry sector classification.

\subsection{Comparing partitions} \label{comp}

Once the stocks are clustered into communities according to their MST links, a comparison with the sectoral communities is performed. In so doing, a measure of similarity of partitions is used to quantify the extent to which the partition, delivered by \textit{Gephi's} Community Detection Algorithm, is close to the sectoral partition. We follow reference (\cite{Fo10}), accounting for the fraction of \emph{correctly classified} nodes in the entire MST.

 We consider two partitions $S =( S_1,S_2,...,S_n)$ and $G = (G_1,G_2,...,G_m)$ of a MST with $n$ and $m$ clusters, respectively, corresponding to the sectoral communities (defined by the classification of the 296 stocks into 11 distinct industry sectors) and to \textit{Gephi's} Community Detection Algorithm (\cite{Bl08}).

A node in the sectoral community $S_i$ is correctly classified, if it gathers in the same \textit{Gephi's} community $G_j$ with at least half of its sectoral partners in $S_i$. This number is divided by the size of the network (296), providing a value between 0 and 1.

Therefore, besides the above mentioned global parameters (modularity, characteristic path length and diameter), the structural aspects of the minimal spanning trees, arising from \emph{business-as-usual}, \emph{Great Recession} and \emph{Global Commodity Crisis} periods, are further described by quantifying the fraction of correctly classified nodes ($\sigma$). In so doing, we aim at contributing to a comprehensible understanding of how community structures change during turbulent periods, by shedding further light on the role of a macroeconomic variable and its influence on stock returns.

In the same way in which a high modularity ($Q^G$) measures a large deviation of the detected clustering from a random network, the fraction of correctly clustered nodes ($\sigma$) provides a complementary quantification of the extent to which clustering conforms to a well known macro structure. Since the communities being defined are exclusively dependent on the strength of the links between stocks, the emerging clusters reflect synchronization in the market. In so doing, its conformity to industry sectors may provide information on the role industry sectors play either in \emph{business-as-usual} or in turbulent periods.

\section{Results} \label{sec:Res}

The underlying data set comprises 296 companies of the \textit{Standard\&Poor's 500} index. Depending on the focus of the next subsections, our analysis and their corresponding results cover either just the pre-crises period (2004-2005), or the turbulent periods of the \emph{Great Recession} (2007-2008) and the \emph{Global Commodity Crisis }(2010-2011).

\subsection{Data overview} \label{over}

The first plot in Figure 1 shows the distribution of the number of companies by industry sector. Companies sort themselves into eleven distinct industry sectors, with \textit{Consumer Discretionary} representing the largest number of companies. The \textit{Telecommunication Services} sector, in contrast, only covers three stocks. The second plot in Figure 1 shows the average values of market capitalization by sector measured in three distinct time periods: before the crises (2004-2005) and in the two turbulent periods of the \emph{Great Recession} (2007-2009) and the \emph{Global Commodity Crisis} (2010-2011). Significant differences between the pre-crises and the crises periods seem to be restricted to the \textit{Energy} and \textit{Telecommunication Services} sectors. In these two sectors, the critical time interval led to the largest absolute increases in market capitalization. Even though the \emph{Great Recession} is predominantly acknowledged as the period of the most severe economic downturn since the \emph{Great Depression} in 1929 (\cite{IMF18}), the \emph{Standard \& Poor's 500 Index} responded only in late 2008 to the turmoil in the banking sector. Thus, the sectors displayed in Figure 1 almost uniformly experienced an increase in average market capitalization between the pre-crisis years and the twenty-four months of 2007 and 2008. With no significant downturn in the S\&P 500 being visible until autumn 2008 and with the stock market not having fully recovered its 2007 peak level until early 2013, most industry sectors still report their average market capitalization throughout the \emph{Global Commodity Crisis} to range below the level of the years 2007 and 2008.

\begin{figure}[h]
\begin{center}
\includegraphics[scale=0.45]{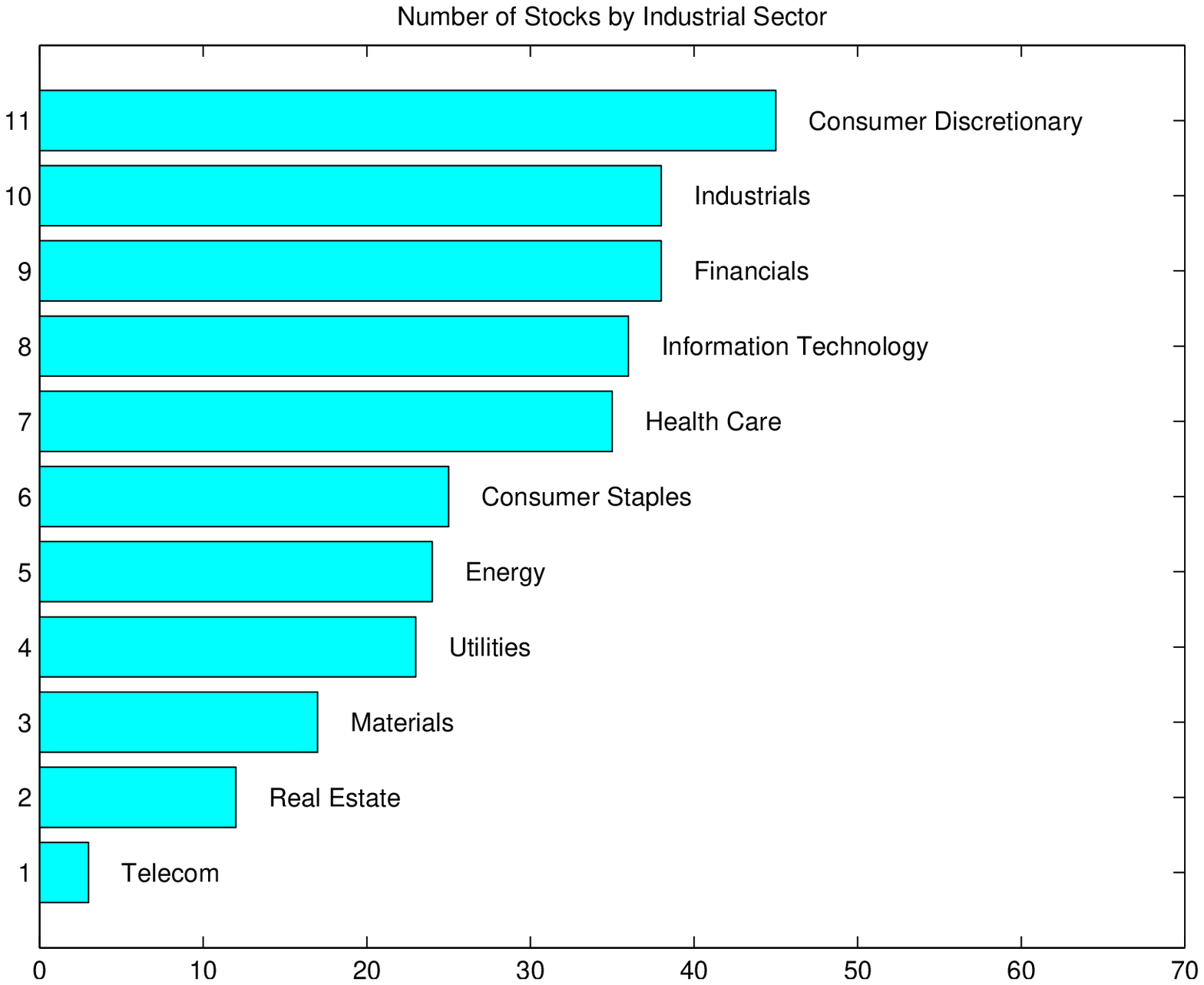}\includegraphics[scale=0.45]{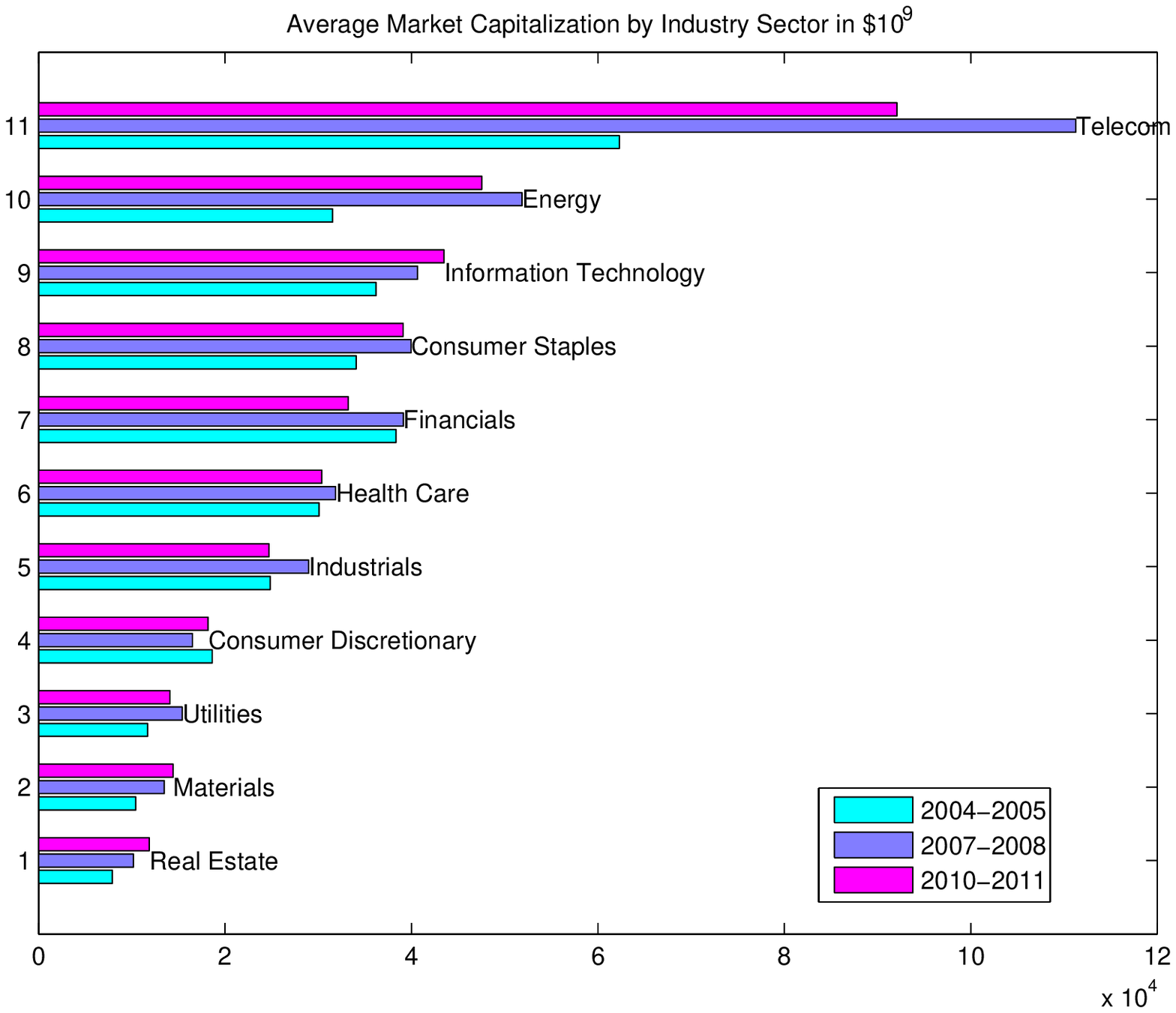}
\caption{The Sectoral distribution of stocks (left plot) and The average Market Capitalization by Industry Sector (right plot) }\label{M01}
\end{center}
\end{figure}

A closer look reveals that only companies of the \emph{Consumer Discretionary} sector could not increase their average market capitalization when comparing the twenty-four months of 2004-2005 and 2007-2008. This picture, however, changes when inspecting the transition from the \emph{Great Recession} to the \emph{Global Commodity Crisis}: only four industry sectors reported an increased average market capitalization in the years 2010 through 2011 with respect to the \emph{Great Recession}, with \emph{Consumer Discretionary}'s relative growth ranking second place, behind the \emph{Real Estate} sector. The largest losses in relative terms were reported by the \textit{Telecommunication Services} sector, followed by \textit{Financials} and \textit{Industrials}.

Comparing the correlation coefficients of stock returns in the pre-crises, in the \emph{Great Recession} and in the \emph{Global Commodity Crisis} periods, reveals an unambiguous picture. Figure 2 demonstrates the effect of the periods characterized by financial distress on the dynamics of one-day returns of stock prices in the S\&P 500. The shift in correlations to higher levels and the consequently overall reduction in the pairwise Euclidean Distances, proxying the simultaneous co-movement of stocks, are an indication of an increased synchronization of stock returns during the \textit{Great Recession} and the \emph{Global Commodity Crisis}.

\begin{figure}[h]
\begin{center}
\includegraphics[scale=0.45]{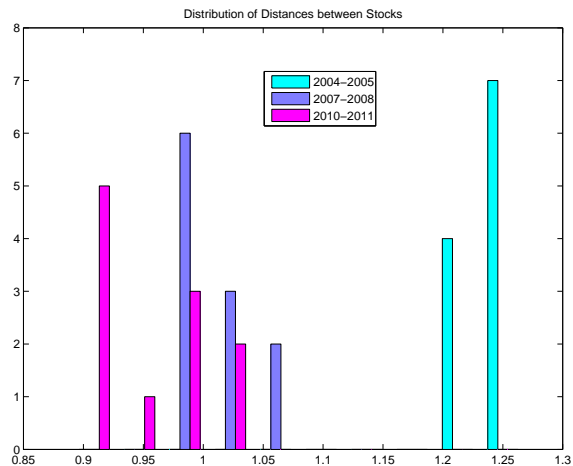}
\caption{Euclidean Distance of Stock Returns}\label{M02}
\end{center}
\end{figure}

\subsection{Network of stocks in the pre-crisis} \label{pre}

After inducting the network of stocks from data of the \emph{business-as-usual} period of 2004-2005 and computing its MST, Figure 3 shows $MST_{04-05}$. There, the size of each node is proportional to the node degree, while nodes are colored according to their industry sector (Industrials: light green; Health Care: orange; Information Technology: black; Utilities: pink; Financials: turquoise; Materials: blue; Consumer Discretionary: purple; Energy: green; Real Estate; yellow; Consumer Staples: red; Telecom: white). The graph layout is generated by \emph{Gephi's} \emph{OpenOrd} algorithm. \footnote{Despite the relatively small number of nodes, \emph{OpenOrd} seemed to better account for the underlying community structure than the \emph{Frutcherman-Reingold} algorithm.}

\begin{figure}[h]
\begin{center}
\includegraphics[scale=0.5]{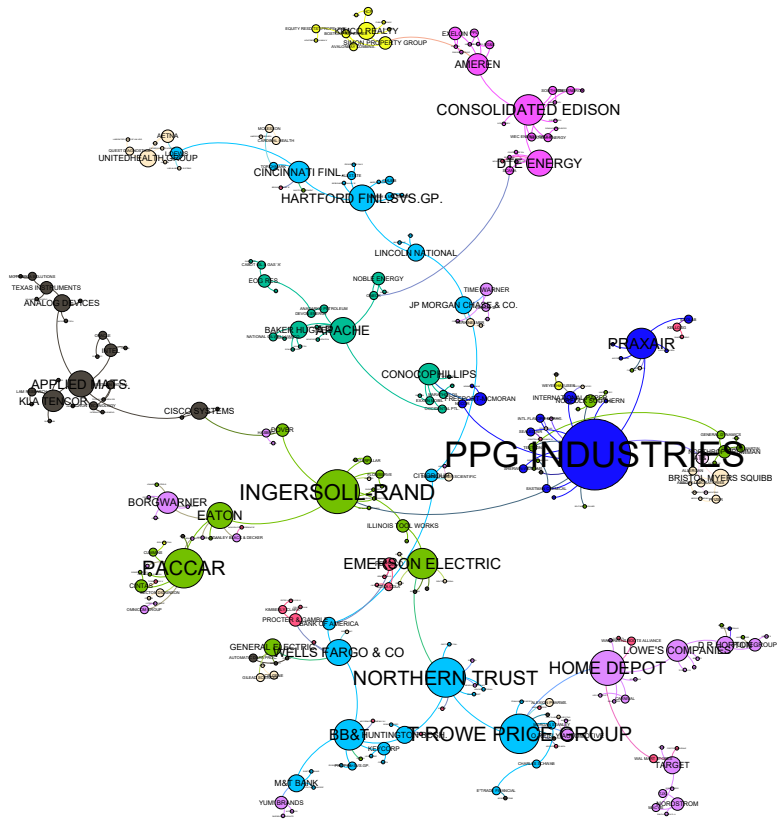}
\caption{The MST from the pre-crisis (2004-2005) data: $MST_{04-05}$}\label{M04}
\end{center}
\end{figure}

Figure 3 shows that \textit{Energy} forms the most connected group of companies, followed by \textit{Financials}. However, the network's hub \textbf{PPG Industries} belongs to the \emph{Materials} sector. With the distribution of the nodes' degree showing some homogeneity, the structure of $MST_{04-05}$ can be considered close to a path motif.

A path motif characterizes a graph in which the number of leaves, i.e., the number of
nodes with degree equal to 1, is much smaller than the size of the graph. A simultaneous consideration of the graph diameter ($d$),
allows to characterize tree motifs with different shapes.

When the number of nodes of the tree is greater than 2,
and depending on the motif that the MST approaches, its diameter ranges in
between $2$ and $N-1$ ($2\le d\leq N-1$). The closer $\frac{d}{N-1}$ is to
1, the less is the similarity of the MST to a \emph{star} motif.
Moreover, with the number of leaves ranging in between exactly the same values but
in the opposite direction, the closer to one, the less is the
similarity of the MST to a \emph{path} motif.

After inducing the network of stocks from the pre-crises data and having computed its MST, the $MST_{04-05}$ diameter ($d$) is calculated and \textit{Gephi's} Community Detection Algorithm is applied to identify communities in $MST_{04-05}$.
Results show that the diameter of $MST_{04-05}$ equals 37, while the characteristic path length ($C$) reaches 11.3. \emph{Gephi's} modularity ($Q^G$) yields 0.88 while the maximum degree ($mk$) is 16.
Computing the fraction of correctly classified nodes ($\sigma_{04-05}$) yields 49, meaning that 49\% of the network's nodes share the same \emph{Gephi's} cluster with their sectoral partners, i.e., those belonging to the same industry sector. Table 1 (Section 4.3) summarizes these results.

As \textbf{PPG Industries} belongs to one of the sectors with the smallest market capitalization, this already suggests, that a clear-cut relationship between a company's size and its interconnectedness, measured in number of degrees, does not exist. Indeed, the 5\% largest companies capture 5.93\% of the minimal spanning tree's links.
A closer look at cluster position and composition in the $MST_{04-05}$ may allow for a comparison with macroeconomic and company-specific circumstances of that time. Whereas the former mostly affects cluster characteristics, individual business features may explain the positioning of certain companies within the network.

Starting with an assessment of common macroeconomic drivers, the two clusters in the upper right-hand corner seem closely linked. Those communities are almost entirely composed of same-sector industries. One is formed by the \emph{Utilities} sector and the other by \emph{Real Estate} companies. According to \cite{GICS} the \emph{Utlities} sector comprises companies working on electricity, gas and water installations, and thus, services required for housing construction. Meanwhile, the MST link between \emph{Real Estate} and the \emph{Utlities} vanish during both the \emph{Great Recession} and the  \emph{Global Commodity Crisis} as displayed in Figures \ref{M05} and \ref{M06}.

Moving on to the individual company assessment, a detail worth examining is the position of the node with the maximum degree (network's hub), \textbf{PPG Industries}. Apparently, the network's second most interconnected company \textbf{Ingersoll Rand}, subsumed under \emph{Industrials}, is located in the fomer's close neighborhood. Indeed, \textbf{Ingersoll Rand} ranges among the five closest companies to \textbf{PPG Industries} in terms of Euclidean distance, whereas only five companies are more similar to \textbf{Ingersoll Rand} than the network's largest hub. In their annual reports of the year 2005 (\cite{PPG05}-\cite{Ing05}), both companies state rising energy and material prices to have increased production costs, but do not name each other as conducting any mutual business relations. Another outstanding node, positioned in the vicinity of \textbf{PPG Industries}, is \textbf{Praxair}. This hub, as well a \emph{Materials} sector company, also stated rising energy prices as having influenced the business development in 2005 (see \cite{Prax05}). This is, nevertheless, not the only link between \textbf{PPG Industries} and \textbf{Praxair}: \textbf{PPG Industries}' chairman of the board and CEO, Raymond W. LeBoeuf, having retired in 2005 after 25 years, is named on the Board of Directors at \textbf{Praxair} during the years 2004 and 2005.

\subsection{Network of stocks in the critical periods} \label{cri}

Different shapes characterize the networks $MST_{07-08}$ and $MST_{10-11}$.

\begin{figure}[h]
\begin{center}
\includegraphics[scale=0.45]{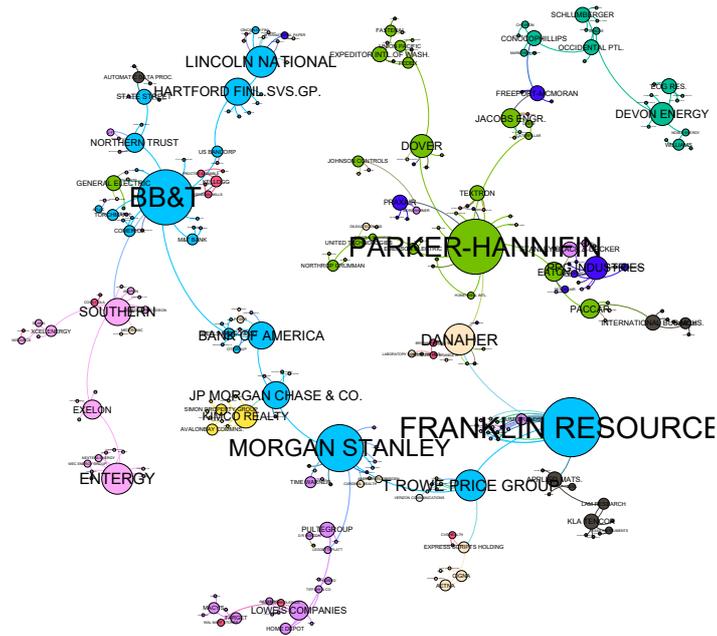}
\caption{The MST from the crisis (2007-2008) data: $MST_{07-08}$}\label{M05}
\end{center}
\end{figure}

\begin{figure}[h]
\begin{center}
\includegraphics[scale=0.5]{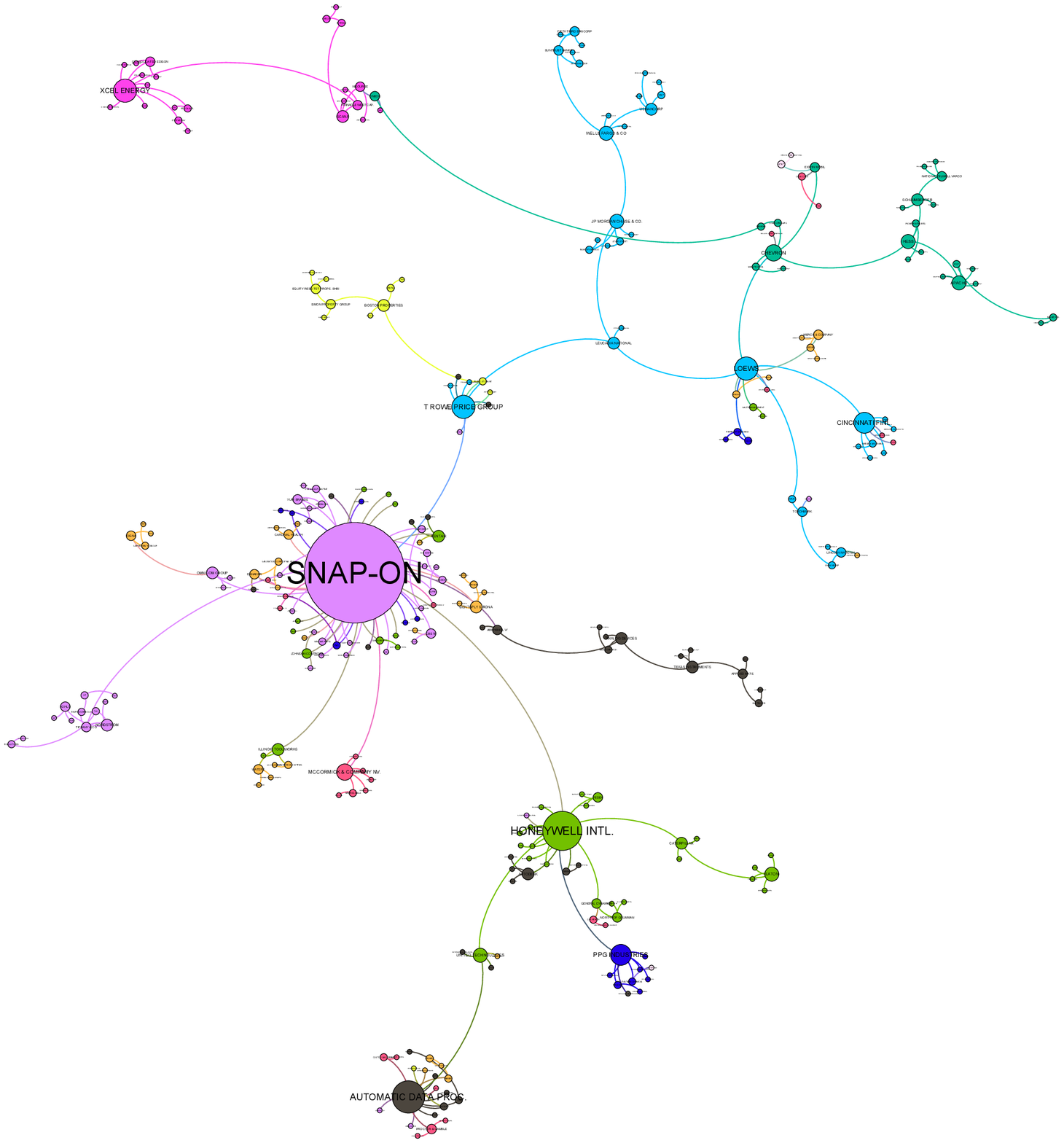}
\caption{The MST from the crisis (2010-2011) data: $MST_{10-11}$}\label{M06}
\end{center}
\end{figure}

Figure \ref{M05} shows the minimal spanning tree ($MST_{07-08}$) induced from the \emph{Great Recession} data. As in Figure 3, the size of each node is proportional to the node's degree, while nodes are colored according to their industry sector, as in Figure 3. After computing the $MST_{07-08}$, \textit{Gephi's} Community Detection Algorithm is applied to identify communities in the $MST_{07-08}$.

Figure \ref{M05} shows that the degree (the size) of some financial companies is much larger than their size in Figure \ref{M04}. The degree distribution reveals that three \emph{Financials} exhibit a degree equal to or larger than 12 and rank among the four most interconnected firms, whereas none of the financial entities reached a degree higher than 9 in the previous assessment period. Furthermore, eight out of the twelve most interconnected nodes are subsumed under the financial industry sector.

\textbf{Franklin Resources} increased its degree from 1 to 15 and \textbf{Morgan Stanley} formed 10 more connections. But also \textbf{BB\&T} doubled its number of links whereas the former period's hub, \emph{PPG Industries} from the \emph{Materials} sector, had to give up 62.5\% of its links. Same-sector agglomeration seems to be limited, with especially the financial hubs being spread out through the system and linking predominantly with extra-sector companies - a hint for the spillover of the banking crisis onto the real economy.

The relationship between degree centrality and size still shows an ambigous picture: the degree representation among the 5\% largest companies could slightly increase to 6.44\%. However, 40\% of the 5\% largest companies showed detrimental dynamics in degree and average market capitalization.

All in all, the overall cluster formation in Figure \ref{M05} represents the partly contradicting market forces. Whereas the banking system had been in trouble since summer 2007 \cite{IMF18}, the S\&P 500 index did not show any sign of bearish behavior until autumn 2008. With the \emph{Real Estate} sector forming a pure cluster in every assessment period, its close attachment to one of the \emph{Financial}'s hubs is clearly visible. A closer look identifies this hub as \textbf{J.P. Morgan Chase}, the investment bank which took over the troubled investmnet bank \textbf{Bear Stearns} in March 2008. \textbf{Bear Stearns} itself was deeply involved in the home mortgage business (see \cite{Hsu10}).

The widespread trouble in the banking sector was at first only limited to the interbank market and recipients of the first rescue measures, such as the \emph{Term Auction Facility}, stayed anonymous (\cite{Posz08}). Still in March 2008, the troubled investment bank \textbf{Bear Stearns} was bailed out with the help of the Federal Reserve \cite{Jen12}, an event which might have prolonged the bullish market sentiment and the neglectance of imbalances in the financial sector. Thus, the \emph{Financials} still stay highly spread out over the network's landscape and interlinked with various industry sectors.

Results on the overall MST characteristics indicate that the diameter of $MST_{07-08}$ equals 29, while the Characteristic Path Length ($C$) equals 10. \emph{Gephi's} modularity, $Q^G$, yields 0.88, as in 2004-2005. These results are summarized in Table 1 (Section 4.3).
The distribution of the size of \emph{Gephi's} clusters is less homogeneous when compared with the results obtained previously.
The calculation of the fraction of correctly classified nodes ($\sigma_{07-08}$) yields 46.1, meaning that 46.1\% of the network's nodes share the same \emph{Gephi's} cluster with those belonging to the same industry sector.

The high value of $Q^G$ during the \emph{Great Recession} is insofar remarkable, as Newman and Girvan (\cite{NewGi04}) mention the usual modularity measure to range between 0.3 and 0.7. Comparing Figures \ref{M04} and \ref{M05}, the predominant characteristic of the latter is the changed degree distribution and appearance of further hubs, especially in the \emph{Financials} sector.

Moving to the analysis of the last two-years period,
Figure \ref{M06} displays the network's landscape during the \emph{Global Commodity Crisis} of the period 2010-2011 ($MST_{10-11}$). Again the picture differs vastly from both the community structure during the \emph{Great Recession} (Figure \ref{M05}) and the pre-crisis period of 2004-2005 (Figure \ref{M04}). The clearly visible hubs of the previous periods vanished almost entirely with only the \emph{Consumer Discretionary} company, \textbf{Snap-On}, forming the new center of attraction. Having even lost a rank since 2007-2008 in terms of average market capitalization, the tools-manufacturer captured 14.6\% of the $MST_{10-11}$ 295 links. Even the second most interconnected firm, \textbf{Honeywell International}, was only assigned 16 links, which is still greater than the 15 links captured by the \emph{Great Recession}'s largest hub.

The overall degree distribution is much more homogenous than during the previous periods: the 5\% most interconnected nodes of the years 2004/2005 captured 20.2\% of the overall links, whereas the ratio increased to 22.5\% during the \emph{Great Recession}. Excluding the largest hub, this ratio fell to 19\% during the \emph{Global Commodity Crisis}.

Cluster formation is again clearly visible, with same-sector clusters being more clear-cut than in previous periods, even though some mix-up still occurs around the network's hubs. In contrast to Figures \ref{M04} and \ref{M05} the \emph{Financials} interact more with same-sector entities and are less linked to other industries. Regarding the largest companies, the \emph{Global Commodity Crisis} seems to have reduced the importance of the largest companies in terms of average market capitalization, as the 5\% largest companies only capture 5.6\% of the $MST_{10-11}$'s links.

Apparently the two hubs of Figure \ref{M06}, \textbf{Snap-on} and \textbf{Honeywell International}, are positioned in a close neighborhood. The type of companies, clustering around the two hubs is heterogenous with regard to industry sectors, however, dominated by \emph{Consumer Discretionary}, \emph{Consumer Staples}, \emph{Industrials} or \emph{Materials}. The common denominator of those sectors is their reliance on commodities. A sub-group of commodities is captured in the \emph{Fuel Index}. A relative dispersion between the \emph{Non-Fuel} and \emph{Fuel} stocks is visible in the $MST_{10-11}$, where the \emph{Energy} and \emph{Utilities} sector are decoupled from those sectors, being less sensitive to energy price fluctuations (see \cite{GICS}). In addition, the obvious same-sector clustering of \emph{Financials} throughout this period coincides with tensions in the Euro Area and widespread doubts about its long-term survival. This hints at the existence of macroeconomic fundamentals in shaping the market's landscape.

The diameter of $MST_{10-11}$ equals 19, while the characteristic path length ($C$) equals 7.1. \emph{Gephi's} modularity, $Q^G$, yields 0.86.
The compliance of \textit{Gephi's} clusters with the \textit{GICS}' sector classification is more pronounced during the \textit{Global Commodity Crisis}: whereas 53.5\% of the sample's companies gathered with their sectoral peers in the same cluster in 2010/2011, this fraction of correctly classified nodes equals only 49\% during the years 2004-2005.

Table 1 summarizes the results presented in this section. The most remarkable outcome is the increase in the maximum degree ($mk$) from 15 to 46 as well as the decrease of the network diameter from 37 to 19. These two aspects drive the shape of the MST in each period. On the contrary, the values of $Q^G$ remain almost unchanged since 2004-2005. Likewise, the small fluctuation of the value of the compliance ($\sigma$) with the industry sector seems to indicate the weak influence of the sectoral classification on the way stocks organize themselves during crises.

\begin{table}[htbp]
  \centering
  \caption{Topological Coefficients of the Minimal Spanning Trees}
    \begin{tabular}{p{2.5cm} p{1.5cm} p{1.5cm} p{1.5cm} p{1.5cm} p{1.5cm}}
    \hline
        & \multicolumn{1}{c}{\emph{d}} & \multicolumn{1}{c}{\emph{C}} & \multicolumn{1}{c}{\textbf{$Q^G$}} & \multicolumn{1}{c}{\textbf{$\sigma$}} & \multicolumn{1}{c}{$mk$} \\
          \cmidrule{2-6}
    $MST_{04-05}$  & \multicolumn{1}{c}{37}    & \multicolumn{1}{c}{11.3} & \multicolumn{1}{c}{0.88} & \multicolumn{1}{c}{49.0} & \multicolumn{1}{c}{16} \\
    $MST_{07-08}$  & \multicolumn{1}{c}{29}    & \multicolumn{1}{c}{10.0} & \multicolumn{1}{c}{0.88} & \multicolumn{1}{c}{46.1} & \multicolumn{1}{c}{15} \\
   $MST_{10-11}$  &   \multicolumn{1}{c}{19}    &  \multicolumn{1}{c}{7.1}     &  \multicolumn{1}{c}{0.86}     & \multicolumn{1}{c}{53.5}  & \multicolumn{1}{c}{43}  \\
    \hline \hline
    \end{tabular}%
  \label{tab:netcharac}%
\end{table}%

\section{Concluding Remarks}

Recalling the purpose of our paper, we were interested in the dynamics of the stock market over time, in particular, in how a changing macroeconomic environment affects the market's inherent community structure. Thus, we took a balanced sample of the S\&P 500, composed of 296 stocks, throughout the period ranging from the beginning of 2004 to the end of 2011, within we contrasted the \textit{business-as-usual} era of 2004-2005 with the subsequent \textit{Great Recession} in 2007-2008 and the \emph{Global Commodity Crisis} of the years 2010-2011.

To uncover the S\&P 500's community structure, we applied a three-fold approach: at first, networks of stocks are induced from a \emph{business-as-usual} (2004-2005), the \textit{Great Recession} and \emph{Global Commodity Crisis} data. To abstract from a fully connected network, we used the Minimal Spanning Tree to filter the strongest links between companies. Based on these shortest distances, the underlying community structure within the three time periods was then characterized by the network's diameter, the characteristic path length, \emph{Gephi's} modularity and the maximum degree. A last step then compared the resulting clusters, produced by \emph{Gephi's} Community Detection Algorithm, with the \emph{natural} partition based on the \emph{GICS'} sector identifier.

The results highlight, how the Euclidean distances among stocks contract in periods of unrest. Furthermore, the steeply decreased values of the network diameter and characteristic path length during both the \textit{Great Recession} and the \emph{Global Commodity Crisis} already indicate a reinforcement of structure within the network. The order of \textit{Gephi's} modularity $Q^G$ reveals a highly clustered system of stocks in the three periods, with same-sector clustering being slightly more pronounced during the years 2010-2011.
In so doing, this study contributes to a further understanding of the time-varying structure underlying the S\&P500, improving the search for {\it economic factors} which may be neither industry sectors nor other obvious economic facts.

\section*{Acknowledgement}

Financial support from national funds by FCT (Funda\c{c}\~{a}o para
a Ci\^{e}ncia e a Tecnologia). This article is part of the Strategic
Project: UID/ECO/00436/2013.

\end{document}